\documentclass[10pt,conference]{IEEEtran}
\IEEEoverridecommandlockouts
\usepackage{cite}
\usepackage{amsmath,amssymb,amsfonts,amsthm}
\usepackage{algorithm,algpseudocode}
\usepackage{graphicx}
\usepackage{booktabs}
\usepackage{svg}
\usepackage{textcomp}
\usepackage{xcolor}
\usepackage{hyperref}
\usepackage{physics}
\usepackage{bm}
\usepackage{array}
\usepackage[normalem]{ulem} 
\usepackage{tikz}
\usepackage{cleveref}
\usepackage{url}
\usetikzlibrary{decorations.pathreplacing}

\DeclareMathOperator*{\argmin}{argmin}
\DeclareMathOperator{\Real}{Re}

\newtheorem{definition}{Definition}

\newcommand{\red}[1]{\textcolor{red}{#1}}
\newcommand{\bfred}[1]{\red{\textbf{#1}}}

\usepackage{subcaption}
\newcommand{\sidecaption}[1]
{\raisebox{\abovecaptionskip}{\begin{subfigure}[t]{1.6em}
			\caption[singlelinecheck=off]{}
			\label{#1}
	\end{subfigure}}\ignorespaces}

\def\BibTeX{{\rm B\kern-.05em{\sc i\kern-.025em b}\kern-.08em
    T\kern-.1667em\lower.7ex\hbox{E}\kern-.125emX}}

\title{Solving Constrained Combinatorial Optimization Problems with Variational Quantum Imaginary Time Evolution}

\begin{document}
\author{\IEEEauthorblockN{Xin Wei LEE}
\IEEEauthorblockA{\textit{School of Computing and Information Systems} \\
\textit{Singapore Management University}\\
80 Stamford Rd, Singapore \\
xwlee@smu.edu.sg}
\and
\IEEEauthorblockN{Hoong Chuin LAU}
\IEEEauthorblockA{\textit{School of Computing and Information Systems} \\
\textit{Singapore Management University}\\
80 Stamford Rd, Singapore \\
hclau@smu.edu.sg}}

\maketitle

\begin{abstract}
    Solving combinatorial optimization problems using variational quantum algorithms (VQAs) has emerged as a promising research direction.
    Since the introduction of the Quantum Approximate Optimization Algorithm (QAOA), numerous variants have been proposed to enhance its performance.
    QAOA was later extended to the Quantum Alternating Operator Ansatz (QAOA+), which generalizes the initial state, phase-separation operator, and mixer to
    address constrained problems without relying on the standard Quadratic Unconstrained Binary Optimization (QUBO) formulation. However, QAOA+ often requires
    additional ancilla qubits and a large number of multi-controlled Toffoli gates to prepare the superposition of feasible states, resulting in deep circuits
    that are challenging for near-term quantum devices. Furthermore, VQAs are generally hindered by issues such as barren plateaus and suboptimal local minima.
    Recently, Quantum Imaginary Time Evolution (QITE), a ground-state preparation algorithm, has been explored as an alternative to QAOA and its variants.
    QITE has demonstrated improved performance in quantum chemistry problems and has been applied to unconstrained combinatorial problems such as Max-Cut.
    In this work, we apply the variational form of QITE (VarQITE) to solve the Multiple Knapsack Problem (MKP), a constrained problem, using a Max-Cut-tailored
    ansatz. To the best of our knowledge, this is the first attempt to address constrained optimization using VarQITE. We show that VarQITE achieves
    significantly lower mean optimality gaps compared to QAOA and other conventional methods.
    Moreover, we demonstrate that scaling the Hamiltonian coefficients can further reduce optimization costs and accelerate convergence.
\end{abstract}

\section{Introduction}
Combinatorial optimization problems (COP) are always of utmost interest among the mathematical optimization community,
as most of the COPs are NP-hard to solve using a classical computer. Variational quantum algorithms (VQA) have emerged as promising tools for solving
COPs since the introduction of the Quantum Approximate Optimization Algorithm (QAOA) by Farhi et al.~\cite{farhi2014quantum}
Intensive studies have been conducted showing the advantage of solving the infamous Max-Cut problem (the problem that divides the graph
into two partitions such that the number of edges between them is maximized) using QAOA.
Some of them showed that QAOA (or its variants) outperforms the Goemans-Williamson algorithm, the best known classical heuristic for Max-Cut,
in terms of the approximation ratio~\cite{xqaoa}.
However, QAOA for constrained problems is not as straightforward as solving Max-Cut, which is an unconstrained problem. For constrained problems,
there are other aspects to consider, e.g. the feasibility of the solution, instead of just the approximation ratio.

To tackle constrained problems, QAOA was extended to the Quantum Alternating Operator Ansatz (QAOA+)~\cite{qaoa-plus}.
QAOA+ allows customized initial states and mixer Hamiltonians that confine the state evolution within the feasible subspace,
avoiding the need for a penalty QUBO formulation.
However, QAOA+ also has its own drawbacks, mainly in the difficulty in creating the superposition of feasible states, and also the complicated
construction of mixers to restrict the transition of the feasible states. These usually result in deep and large circuits that are impractical
for noisy quantum devices~\cite{grover-mixer,xie2024feasibility,gm-ub,qaoa-review}.

On the other hand, expressive ansatz circuits like the hardware-efficient ansatz (HEA)~\cite{hea},
multi-angle QAOA (ma-QAOA)~\cite{ma-qaoa}, and expressive QAOA (XQAOA)~\cite{xqaoa} can also be used to solve Max-Cut and beyond.
However, the expressivity of the ansatz is also strongly tied to the existence of barren plateaus in the optimization landscape~\cite{McClean2018,Fontana_2024,Ragone_2024}.
Barren plateau is a phenomenon that commonly and inherently occurs in deep and expressive quantum circuits, where the gradients
in a particular region of the landscape are near to zero. As the problem size scales up, it will be more difficult to get a good solution using expressive ansatzes,
along with more occurrences of local (sub-optimal) minima.
Having many local minima is also another problem that haunts the optimization of VQA. Due to the oscillatory nature of parameterized unitary gates
(consists of sines and cosines), minimization of the Hamiltonian expectation using gradient-based methods is bound to be non-convex.
Therefore, heuristics are proposed to determine good parameter initializations, especially for QAOA which exhibits certain patterns in
parameters~\cite{leapfrog,warm-start,galda2021transferability,tqa2021,lee2021parameters,bilinear}.

Other efforts made to improve the quality of the solution for constrained COPs include tampering with the formulation of the problem~\cite{space-efficient},
using different ansatz circuits~\cite{xy-mixer,lcc}, different optimization heuristics~\cite{lw-biamonte,itlw}, etc. 

Recently, Quantum Imaginary Time Evolution (QITE) has been explored as a possibility to improve performance in solving constrained COPs.
In contrast with the quantum time evolution which is used to simulate the evolution of a given Hamiltonian with time, QITE is used to evolve the system
such that it converges to the ground state of the given Hamiltonian. 
The main challenge of QITE lies in the implementation of the non-unitary imaginary time operator $e^{-\tau\hat{H}}$ in a gate-based quantum computer that only
allows unitary quantum gates for state evolution.
There are mainly two main paradigms for implementing QITE: 1) the non-variational QITE~\cite{qite-seminal} that breaks $e^{-\tau\hat{H}}$ into fragments using
Trotterization to mimic the imaginary time evolution for each small step $\Delta\tau$,
and 2) the variational QITE (VarQITE)~\cite{var-qite} that utilizes the McLachlan variational principle to update the parameters in the ansatz circuit~\cite{McLachlan}.
In general, non-variational QITE requires exponentially many gates to be implemented, which is not NISQ-friendly~\cite{qite-nonlocal}.
Therefore, we will focus on the variational QITE.

There are several works on solving COPs using the non-variational QITE framework.
Alam et al. proposed a linear ansatz, which essentially contains only single-qubit rotational-$Y$ gates, to solve the Max-Cut problem using QITE~\cite{philip-qite}.
Due to the lack of entangling gates in the ansatz circuit, it can be efficiently simulated on a classical computer.
They solved Max-Cut of up to 50 nodes and obtained an average approximation ratio of 0.89 with QITE on linear ansatz. 
By applying optimization tricks, the average approximation ratio is increased to 0.97. 
However, the authors did not compare their method with competitive methods, such as QAOA or the Goemans-Williamson (GW) algorithm for Max-Cut.
Instead, they compare their results with the classical greedy algorithm. 

Bauer et al. extended the previous work to solve the Low Autocorrelation Binary Sequence (LABS) problem using QITE with linear ansatz~\cite{pubo-qite}.
The LABS problem is an unconstrained, higher-order problem compared to Max-Cut. First, the authors compared QITE and the classical GW in solving weighted Max-Cut for up to 150 qubits. 
In some of the cases, QITE shows better approximation ratios than GW, but GW still outperforms QITE on average. 
For LABS, the authors solved the problems up to 28 qubits, the results for QITE are on a par with that for QAOA with depth-10. 

On the other hand, Wang et al. recently proposed the imaginary Hamiltonian variational ansatz (iHVA) to solve the Max-Cut problem~\cite{symmetry-enhanced,ihva}.
This ansatz is inspired by QITE, and is designed based on the bit-flip symmetry of the Max-Cut problem
(if $b$ is the solution bit-string of the Max-Cut problem, then the bit-string $b'$ obtained by flipping all the bits in $b$ is also the solution). 
The authors solved the Max-Cut problem using iHVA, with the variational parameters of the ansatz optimized by a classical optimizer.
Despite being QITE-inspired, the authors did not use the QITE algorithm to solve Max-Cut. Therefore, we are going to solve our target problem using iHVA combined with QITE.

In this work, we focus on the Multiple Knapsack Problem (MKP), a problem with multiple inequality constraints. MKP has important applications in logistics, resource allocation and scheduling. Knapsack-like problems are also studied by many using
VQAs~\cite{delagrandrive2019knapsack,qo-heuristics-knapsack,qalg-knapsack,qc-technique-mkp,monit-mkp}.
Contrary to the standard 0-1 Knapsack problem which is fairly easy to solve by classical computers in pseudo-polynomial time in the worst case using dynamic programming, the MKP has no pseudo-polynomial time algorithm. Furthermore, whereas the standard 0-1 Knapsack has a fully polynomial-time approximation scheme (FPTAS),
MKP in general is considered hard as no FPTAS exists, unless $\mathsf{P=NP}$.

MKP is converted to a Quadratic Unconstrained Binary Optimization (QUBO) problem,
which is an unconstrained formulation that penalizes the infeasible solutions. We use the unbalanced QUBO approach proposed in~\cite{ub-penalty} to omit the use of the slack variables which will introduce
extra qubits which will be challenging to scale. We compare the solution quality of different methods: iHVA with QITE, iHVA without QITE, and the multi-angle QAOA (ma-QAOA).
Since iHVA is tailored for Max-Cut, our QUBO-formulated problems are converted to the Max-Cut problems, then they are converted to their corresponding iHVA circuits.

We show that in solving our MKP instances, QITE shows significantly lower mean optimality gap compared to conventional optimization methods.
Moreover, we also show that by scaling the Hamiltonian coefficients, QITE can achieve better performance in terms of finding the optimal solution, as
well as faster convergence with a smaller number of time steps.

\section{Background}
\subsection{Problem formulation}
Given $m$ knapsacks with limited capacities $W_i$ and $n$ items with respective values $v_j$ and weights $w_j$,
MKP seeks to assign items to the knapsacks such that the combination maximizes the value of the items carried,
such that the weights of the items $w_j$ do not exceed the capacity of the knapsack $W_i$, and each knapsack can contain only one item.
The binary decision variables are denoted as $x_{ij}\in\{0,1\}$, where $x_{ij}=1$ if item $j$ is placed in knapsack $i$ (and 0 otherwise).
It is formally defined as
\begin{align}
    \max_{\mathbf{x}} & \sum_{i=1}^m \sum_{j=1}^n v_jx_{ij} \\
    \text{s.t.} & \sum_{j=1}^n w_jx_{ij} \leq W_i,\quad i = 1,...,m \\
    & \sum_{i=1}^m x_{ij}\leq 1,\quad j = 1,...,n.
\end{align}
where \\
\begin{tabular}{lll}
    $x_{ij}$& : & Decision variable to represent whether item $j$ is in \\
    & & knapsack $i$. \\
    $v_j$& : & Value for item $j$. \\
    $w_j$& : & Weight of item $j$. \\
    $W_i$& : & Capacity of knapsack $i$.
\end{tabular} \\


Since MKP has inequality constraints, a common way to formulate it as a QUBO is to use slack variables to convert the inequalities to equalities,
then penalty multipliers are added to the objective function to penalize the infeasible solutions. 
However, introducing slack variables is known to be a bad practice in variational quantum optimization.
Besides of introducing an extra number of variables for the problem (hence an extra number of qubits), 
slack variables also deteriorate the optimization landscape of variational algorithms by not reflecting the true objective values for
the solutions in the original problem.  
Therefore, we use the recently proposed unbalanced penalization method~\cite{ub-dwave,ub-penalty} that does not introduce slack variables for inequality constraints.

The unbalanced penalization method follows the intuition to penalize the inequality constraint function $h(\mathbf{x})$ using the exponential function
$e^{-h(\mathbf{x})}$, such that the penalty approaches 0 when $h(\mathbf{x}) > 0$ and is exponentially large when $h(\mathbf{x}) < 0$. In our case, 
the constraint functions are
\begin{align}
    h_1^{(i)}(\mathbf{x}) & = W_i - \sum_{j=1}^n w_jx_{ij} \geq 0,\quad i = 1,...,m \\
    h_2^{(j)}(\mathbf{x}) & = 1 - \sum_{i=1}^m x_{ij} \geq 0, \quad j = 1,...,n.
\end{align}
Note that $e^{-h(\mathbf{x})}$ is added to the objective function, but it is difficult to implement the exponential function as quantum observables.
Therefore, the exponential function is approximated to the second order with the Taylor series expansion:
$e^{-h(\mathbf{x})} \approx 1 - \lambda_1 h(\mathbf{x}) + \lambda_2 h(\mathbf{x})^2$, so the overall objective function is still quadratic.
Constant 1 can be ignored as it does not affect the maximization. The objective function for the unbalanced MKP then becomes 
\begin{multline}
    \min_{\mathbf{x}} -\sum_{i=1}^m \sum_{j=1}^n v_jx_{ij} - \lambda_1 \left[ \sum_{i=1}^m h_1^{(i)}(\mathbf{x}) + \sum_{j=1}^n h_2^{(j)}(\mathbf{x}) \right] \\
    + \lambda_2 \left[ \sum_{i=1}^m h_1^{(i)}(\mathbf{x})^2 + \sum_{j=1}^n h_2^{(j)}(\mathbf{x})^2 \right].
    \label{eqn:final-qubo}
\end{multline}
The maximization problem is converted to a minimization problem by negating the objective term.
Note that the objective function is quadratic in terms of the decision variable. 

To adapt the problem to a Max-Cut tailored ansatz, we follow the algorithm stated in~\cite{qubo2maxcut} to convert our QUBO problem in Eq.~(\ref{eqn:final-qubo})
to a Max-Cut problem. This is useful since Max-Cut is a problem that is intensively studied in quantum optimization. The algorithm states that any QUBO instance
with $n$ variables can be converted to a Max-Cut instance with $n+1$ vertices. Refer to Algorithm~\ref{alg:qubo-to-maxcut} in 
the Appendix for the details of the conversion algorithm.

The QUBO objective function in~(\ref{eqn:final-qubo}) is then converted to the problem Hamiltonian $\hat{H}$ using a linear mapping
\begin{equation}
    x_i = \frac{1-z_i}{2}
\end{equation}
to convert the binary variables $x_i\in\{0,1\}$ to the spin variables $z_i\in\{1,-1\}$.

\subsection{iHVA}
iHVA is a QITE-inspired ansatz that was recently proposed in~\cite{ihva}, tailored to solve the Max-Cut problem.
The ansatz is designed based on the time-reversal symmetry in the Ising Hamiltonian and the bit-flip symmetry in Max-Cut, so that it mimics the
imaginary time evolution that leads the system to the ground state. The resulting ansatz is a series of $R_{ZY}$ gates acting on two qubits:
\begin{align}
    R_{ZY}(\theta)_{i,j} & = e^{-i\theta Z_iY_j/2} \\
    & = \sqrt{X}_j R_{ZZ}(\theta)_{i,j}\sqrt{X}_j.
\end{align}
The second line is for the convenience of implementation. $\sqrt{X}_i$ is the square root $X$ gate ($\sqrt{X}\sqrt{X}=X$) acting on qubit $i$,
and $R_{ZZ}(\theta)_{i,j} = e^{-i\theta Z_iZ_j/2}$ is a common gate used in QAOA.

The iHVA circuit is constructed using the following procedure:
\begin{enumerate}
    \item Find the breadth-first spanning tree in the Max-Cut graph $G$.
    \item Append the circuit with the $R_{ZY}$ gates on the corresponding qubits for each of the edges in the spanning tree.
    \item Remove those edges from $G$ and repeat from Step 1 until no edges are left in $G$.
\end{enumerate}
When repeating the layers of the ansatz, it is encouraged to place $R_{ZY}$ and $R_{YZ}$ alternatively to increase the expressivity of the ansatz. 
As such, the ansatz will have a total of $p|E|$ parameters to be optimized, where $p$ is the number of rounds and $|E|$ is the number of edges in $G$.

\begin{figure*}
    \centering
    \includegraphics[width=1.0\linewidth]{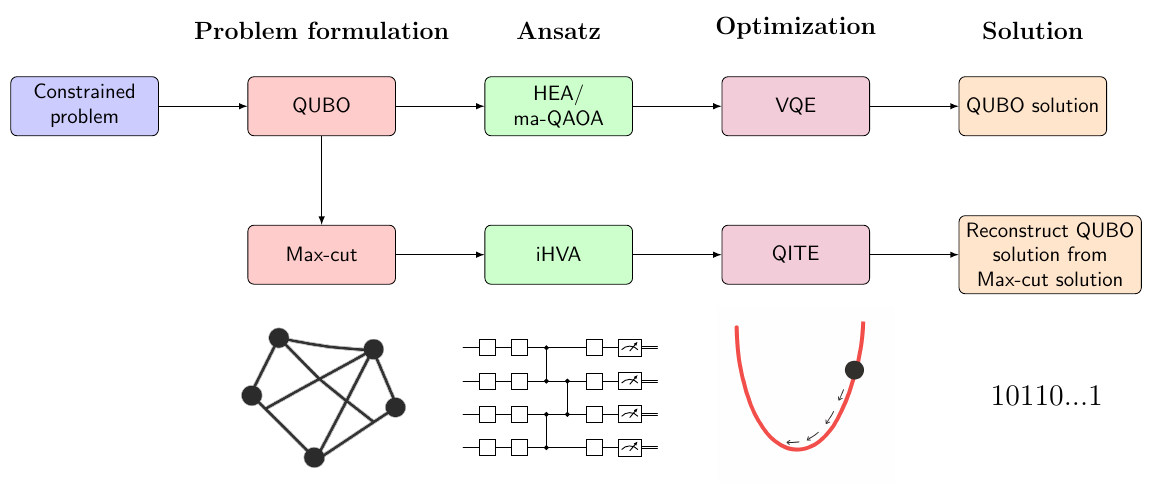}
    \caption{Workflow for the framework used in this work.}
    \label{fig:flowchart}
\end{figure*}

\subsection{Variational QITE}
The core idea of QITE is to replace the real time evolution with ``imaginary time''
$\tau = it$, so the evolution operator becomes $e^{-\tau\hat{H}}$. This evolution is non-unitary, 
so the quantum state need to be normalized after the evolution:
\begin{equation}
    \ket{\Psi(\tau)} = \frac{e^{-\tau\hat{H}}\ket{\Psi(0)}}{||e^{-\tau\hat{H}}\ket{\Psi(0)}||}.
    \label{eqn:qite}
\end{equation}

The challenge of QITE lies in the implementation of the non-unitary operator $e^{-\tau\hat{H}}$. 
For non-variational QITE, the operator is split into fragments with small time steps $\Delta\tau$.
Then, a unitary $e^{-i\Delta\tau\hat{A}[m]}$ is found such that it approximates the imaginary time evolution of
the local fragment of the Hamiltonian $e^{-\Delta\tau h[m]}$, where $\hat{H} = \sum_m h[m]$.
However, doing this requires an exponential number of gates as the time step progresses, as the unitary will need to
involve more and more qubits~\cite{qite-seminal}. Although efforts are made to reduce the number of gates required, 
it still scales exponentially with the locality of the Hamiltonian~\cite{qite-nonlocal}.

To work around the expensive implementation of the non-unitary, a variational version of QITE (VarQITE) has been proposed \cite{var-qite}.
Recall that we want to evolve the quantum state as stated in Eq.~(\ref{eqn:qite}). In VarQITE, assume that we have a unitary
$U(\bm{\theta})$ that is expressive enough to represent the desired state $\ket{\psi(\bm{\theta})}\approx\ket{\Psi(\tau)}$:
\begin{equation}
    \ket{\psi(\bm{\theta})} = U(\bm{\theta})\ket{\psi_0},
\end{equation}
where $\bm{\theta} = \bm{\theta}(\tau)$ is a list of variational parameters to be trained.
The \emph{McLachlan variational principle}~\cite{McLachlan} is used to simulate the evolution of $\bm{\theta}(\tau)$,
which essentially finds the variational parameters that minimizes the difference between both sides of the Wick-rotated Schro\"dinger Equation:
\begin{equation}
    \dot{\bm{\theta}} = \argmin_{\bm{\theta}} \left\lVert\left( \frac{d}{d\tau} + \hat{H} - E_\tau\right)\ket{\psi(\bm{\theta})} \right\rVert,
\end{equation}
where $E_\tau = \expval{\hat{H}}{\psi(\bm{\theta})}$ arises from normalization.
This is equivalent to solving the system of linear equations:
\begin{equation}
    \sum_j M_{ij}\dot{\theta}_j = V_i,
    \label{eqn:lse}
\end{equation}
where
\begin{align}
    M_{ij} & = \Real \left[ \frac{\partial \bra{\psi(\bm{\theta})}}{\partial \theta_i} \frac{\partial 
    \ket{\psi(\bm{\theta})}}{\partial \theta_j} \right], \\
    V_i & = -\Real \left[ \frac{\partial \bra{\psi(\bm{\theta})}}{\partial \theta_i} H \ket{\psi(\bm{\theta})} \right]. \label{eqn:vi}
\end{align}
Theoretically, both $M$ and $V$ can be found efficiently using the Hadamard test as explained in~\cite{var-qite}.
Eq.~(\ref{eqn:lse}) can then be solved by
\begin{equation}
    \dot{\bm{\theta}} = M^{-1}(\bm{\theta})V(\bm{\theta}),
    \label{eqn:solving-lse}
\end{equation}
which are essentially differential equations.
Therefore, the parameters are updated using the Euler method:
\begin{equation}
    \bm{\theta}(\tau_0 + \Delta\tau) = \bm{\theta}(\tau_0) + \dot{\bm{\theta}}\Delta\tau,
    \label{eqn:euler}
\end{equation}
where $\Delta\tau$ is a small interval. For convenience, since we only consider VarQITE in our experiments, we will use QITE to refer to the VarQITE and
explicitly state the ``non-variational'' term when we discuss the non-variational QITE.

\subsection{Entire framework}
To solve constrained problems using variational algorithms, we need to first convert the problem into a QUBO. 
Here, we use the unbalanced formulation of QUBO as discussed in previous sections. Since iHVA is tailored for the Max-Cut problem,
we convert the QUBO to a Max-Cut instance to construct the ansatz. For other ansatzes, the QUBO is solved directly.
Next, we use two different algorithms to optimize the QUBO or Max-Cut loss function: QITE or VQE. QITE is implemented using
the variational McLachlan principle to optimize the trial state produced by iHVA. For VQE, any classical optimizers,
e.g., Broyden-Fletcher-Goldfarb-Shanno (BFGS)~\cite{l-bfgs-b},
Constrained Optimization BY Linear Approximation (COBYLA)~\cite{cobyla}, or Stochastic Gradient Descent (SGD)~\cite{sgd}, can be used for optimization. 
Lastly, the solution is retrieved by sampling the optimal quantum circuit (the parameterized quantum circuit with optimal parameters). The bit-string
with the highest probability is chosen as the solution. For the cases in which the QUBO is converted to a Max-Cut, the Max-Cut solution needs to converted back to the QUBO solution using 
the procedure stated in the Appendix.
The overall procedure to solve the constrained problem using this framework is as follows.
\begin{enumerate}
    \item Convert the constrained problem to an unbalanced QUBO.
    \begin{enumerate}
    \item For iHVA, convert the QUBO to a Max-Cut instance.
    \end{enumerate}
    \item Construct the ansatz based on the QUBO or Max-Cut instance.
    \item Run QITE or VQE to find the ground state of the Hamiltonian.
\end{enumerate}
Fig.~\ref{fig:flowchart} shows the entire workflow for the framework used.

\begin{figure}
    \centering
    \includegraphics[width=0.9\linewidth]{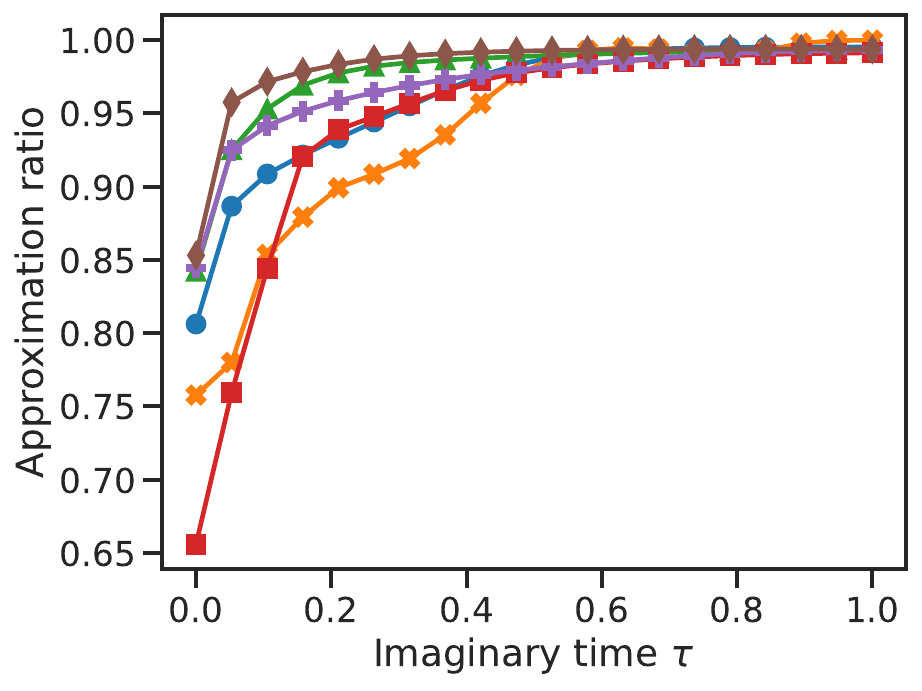}
    \caption{The convergence of QITE+iHVA to the solution of MKP (corresponding Max-Cut) as the imaginary time $\tau$ passes. Each line in the plot represents one instance of MKP.}
    \label{fig:qite-individual}
\end{figure}

\section{Experimental Settings and Results}
We solve MKP with at most 3 knapsacks and 4 items using VarQITE. After formulating as an unbalanced QUBO,
we require 9 to 12 qubits to encode the problem. In the case of using iHVA as the ansatz, the unbalanced QUBO is then converted
to a Max-Cut with one extra qubit (10 to 13 qubits). Fig.~\ref{fig:qite-individual} shows the convergence of QITE with the iHVA ansatz for
6 instances of MKP. Each line in the plot represents the convergence of one MKP instance. The MKP QUBO is converted to the Max-Cut problem to construct the corresponding iHVA ansatz. 
It shows that the approximation ratio increases as the imaginary time $\tau$ passes, showing the capability of QITE to produce a near-optimal
solution for the MKP instances.

We mainly compare the performances of the problems solved using QITE and that without using QITE,
using the iHVA ansatz. We also compare the results with those produced using typical ansatz such as the multi-angle QAOA (ma-QAOA)~\cite{ma-qaoa}
and the Hardware Efficient Ansatz (HEA), optimized using a classical optimizer. The details for the experimental settings are listed as follow: 
\begin{enumerate}
    \item \textbf{Dataset:} 68 instances of Multiple Knapsack Problem, with at most 3 knapsacks and 4 items. The solutions of the instances are non-trivial, i.e., all the instances have at least one item
    in one of the knapsacks.
    \item \textbf{Simulator:} State vector simulation using Qiskit~\cite{qiskit2024}. Probabilities for optimal circuits are sampled using the Qiskit Sampler.
    \item \textbf{Problem formulation:} Unbalanced penalization of QUBO with $\lambda_1 = \lambda_2 = 10$. For iHVA, the unbalanced QUBO is converted to a Max-Cut instance before constructing the ansatz.
    \item \textbf{Initialization:} 5 different random initializations (trials) for every method.
    \item \textbf{Methods:} iHVA optimized using QITE (QITE+iHVA); iHVA, ma-QAOA, and HEA ansatzes optimized using the Broyden-Fletcher-Goldfarb-Shanno (BFGS) algorithm.
    The hyperparameters are set as the default parameters provided in the Qiskit interface (\verb+maxfev=15000+, \verb+maxiter=15000+, \verb+ftol=2.22e-15+)~\cite{bfgs-qiskit}.
    All ansatzes are of one repetition.
    \item \textbf{Evaluation metrics:} Feasibility rate, optimality rate, and mean optimality gap (see definitions).
\end{enumerate}
The solution bit-string is obtained by sampling the optimal circuit (the ansatz circuit substituted with optimal parameters). The bit-string with the highest probability in the sampled probability distribution
is considered to be the solution to the problem. The solution is then substituted back into the constrained program to evaluate its feasibility and optimality. The following metrics are calculated based on the sampled solutions.
\begin{definition}[Feasibility rate]
    The rate of feasible solutions obtained out of a certain number of trials. Feasible solutions are solutions that satisfy all the constraints in the problem.
\end{definition}
\begin{definition}[Optimality rate]
    The rate of optimal solutions obtained out of a certain number of trials. Optimal solutions are solutions that give the minimum objective in
    Eq.~(\ref{eqn:final-qubo}) and are feasible.
\end{definition}
\begin{definition}[Mean optimality gap]
    The average optimality gap out of all trials. The optimality gap is defined as
    \begin{equation}
        \text{opt. gap} = 1 -\frac{C_{\text{VQA}}}{C_{\text{opt}}},
    \end{equation}
    where $C_{\text{VQA}}$ is the objective of the solution given by different VQAs, and $C_{\text{opt}}$ is the objective of the optimal solution.
    The objectives are obtained by substituting the solution bit-string back into the original problem.
    The optimal solution is found by optimizing classically using Gurobi~\cite{gurobi}.
\end{definition}

The feasibility/optimality rate quantifies the percentage of feasible/optimal solutions obtained from the VQA out of a number of random initializations.
We consider a good VQA should give more feasible/optimal solutions with different random initializations.
The mean optimality gap quantifies, on average, how near the solutions obtained using quantum optimization methods are to the optimal solution. Nearer values to zero
means the solution of VQA is nearer to the optimal solution.

\begin{figure}
    \centering
	\begin{subfigure}[t]{0.45\textwidth}
		\centering
		\sidecaption{subfig:a}
		\raisebox{-\height}{\includegraphics[width=\textwidth]{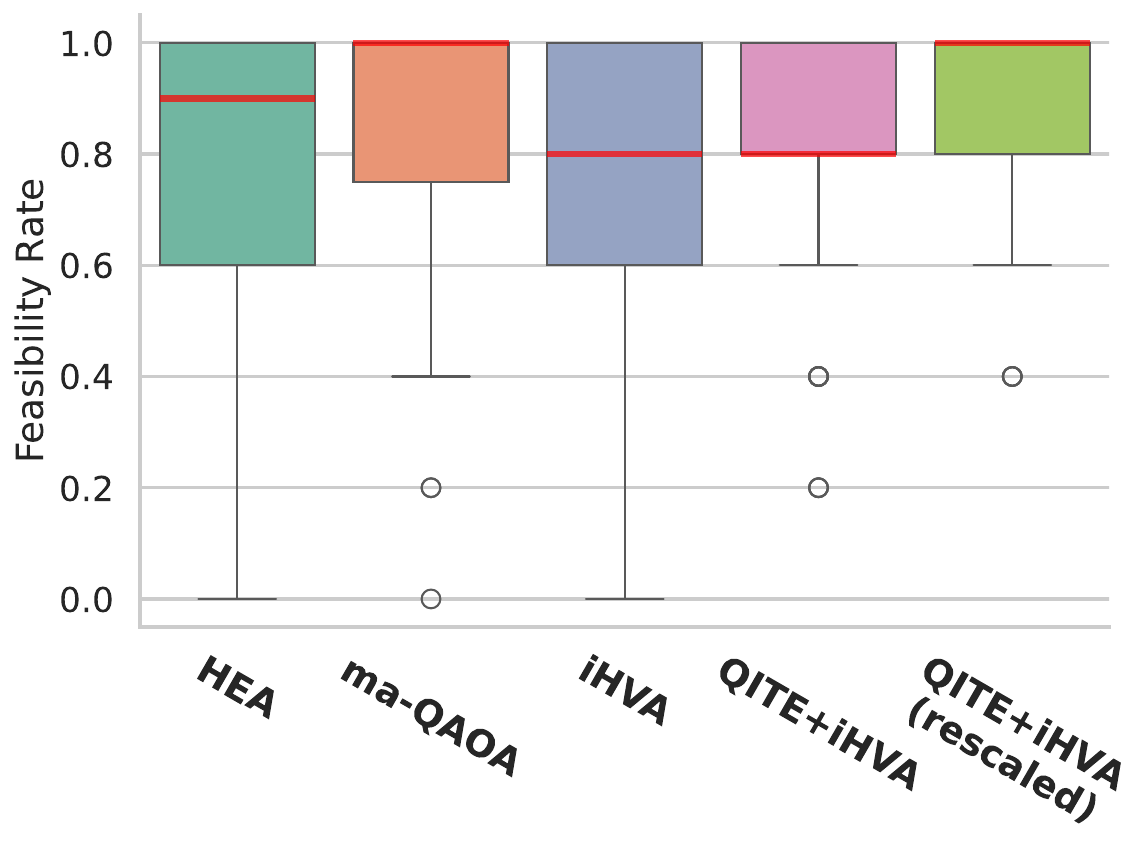}}
	\end{subfigure}
    \begin{subfigure}[t]{0.45\textwidth}
		\centering
		\sidecaption{subfig:b}
		\raisebox{-\height}{\includegraphics[width=\textwidth]{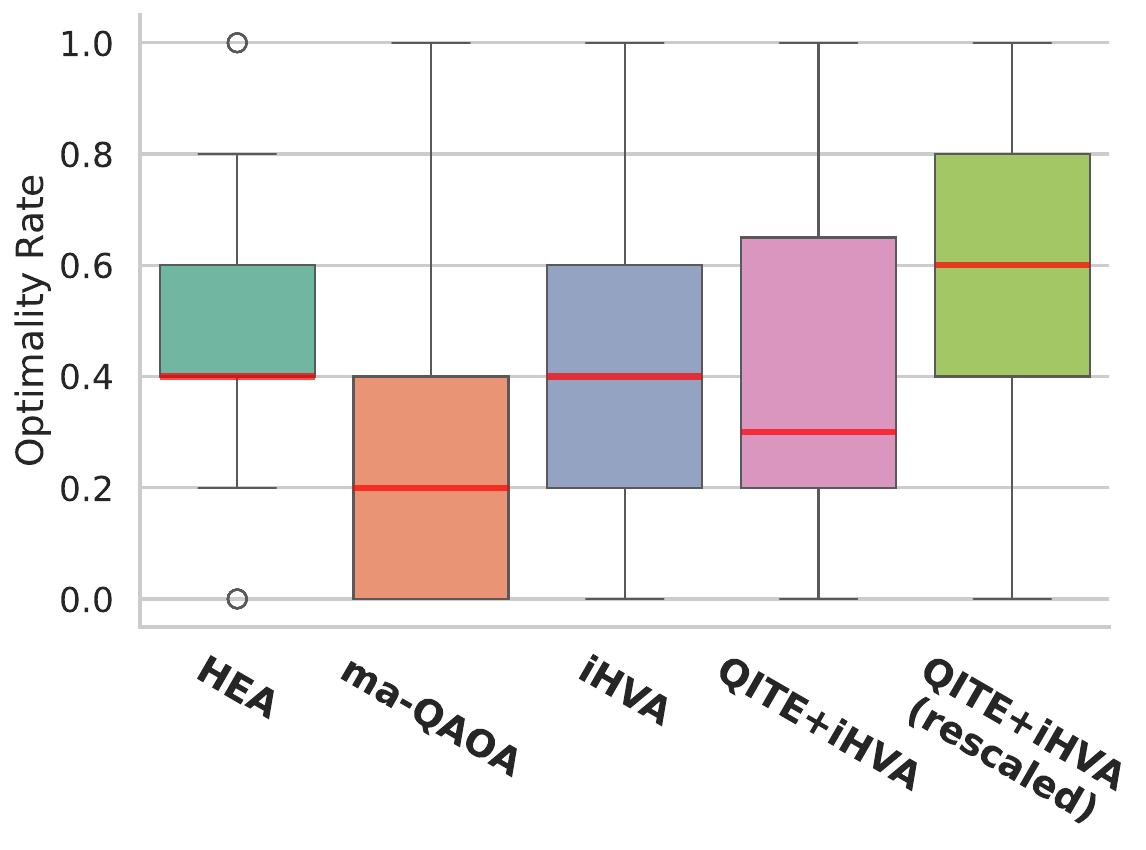}}
	\end{subfigure}
    \begin{subfigure}[t]{0.45\textwidth}
		\centering
		\sidecaption{subfig:c}
		\raisebox{-\height}{\includegraphics[width=\textwidth]{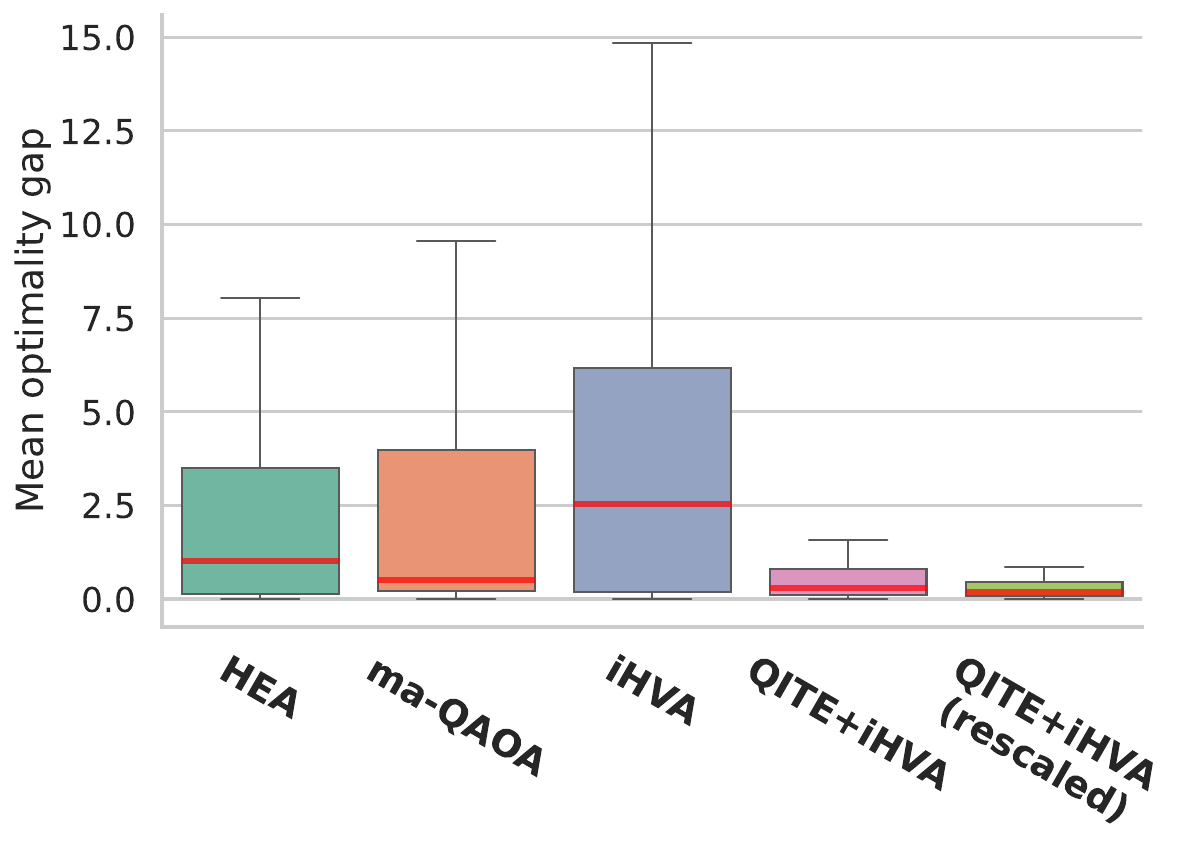}}
	\end{subfigure}
    \caption{Performance comparison of different methods (HEA, ma-QAOA, iHVA, QITE+iHVA, and rescaled QITE+iHVA) across three evaluation metrics: (a) feasibility rate, (b) optimality rate, and (c) mean optimality gap.
    Each box of the boxplot shows the corresponding metric for 68 different MKP instances. The red line in the box shows the median of the corresponding metric.}
    \label{fig:results}
\end{figure}

Fig.~\ref{fig:results}(a)--(c) shows the results for the different methods used to solve MKP. For QITE+iHVA, we also consider a ``rescaled'' version of the Hamiltonian coefficients to
aid the convergence of QITE. We further elaborate on this in the discussions. Fig.~\ref{fig:results}(a) shows the feasibility rate of the methods used.
All methods demonstrate relatively high performance, with the rescaled QITE+iHVA achieving the highest median feasibility rate of 1.0, followed closely by ma-QAOA, also with a median of 1.0, although with a wider spread and some
outliers dropping below 0.2. QITE+iHVA and iHVA both have a median feasibility around 0.85–-0.9, while HEA shows the lowest median of approximately 0.75, along with a broader distribution and several low outliers.

In Fig.~\ref{fig:results}(b), which evaluates the optimality rate, the performance differences become more pronounced. The rescaled QITE+iHVA again leads with a median optimality rate of about 0.65,
followed by QITE+iHVA at around 0.35.
HEA and iHVA show similar median values of roughly 0.4, but with wider interquartile ranges and lower minimum values, especially in iHVA.
ma-QAOA performs the worst, with a median optimality rate near 0.2 and a lower quartile reaching 0, indicating frequent failure to find optimal solutions.

Fig.~\ref{fig:results}(c) focuses on the mean optimality gap, where lower values are better. The rescaled QITE+iHVA outperforms all other methods, showing the smallest median gap of approximately 0.2,
closely followed by QITE+iHVA with a median gap of around 0.5. HEA and ma-QAOA show moderate performance, with median gaps around 2.0–-2.5, while iHVA performs the worst, with a median gap close to 4.0 and a maximum reaching 15,
reflecting high variability and less reliable performance.
It can be seen that although the plots of the feasibility rate and the optimality rate do not differ much in different methods, the mean optimality gap of QITE
is drastically lower than those of the classical VQE methods. This means that the solution given by QITE is on average nearer to the optimal solution than
using classical optimizers.
Overall, these results suggest that the rescaled QITE+iHVA consistently achieves high feasibility and optimality rates with the smallest optimality
gaps, making it the most effective and robust method among those evaluated.

\begin{table*}
    \centering
    \caption{The comparison of QITE methods and VQE methods by different metrics. The feasibility and optimality columns shows the percentage of instances %
    that are feasible/optimal using different methods. An instance is recorded as feasible/optimal if at least one out of 5 random trials is feasible/optimal. %
    The mean feasibility rate, mean optimality rate, and mean optimality gap are the means taken across all 68 instances.}
    \begin{tabular}{cccccc}
        \toprule
        \textbf{Method} & \textbf{Feasibility} & \textbf{Optimality} & \textbf{Mean feasibility rate} & \textbf{Mean optimality rate}
        & \textbf{Mean optimality gap} \\
        & \textbf{(best out of 5)} & \textbf{(best out of 5)} &&& \\
        \midrule
        \textbf{QITE+iHVA (rescaled)} & \bfred{100.0\%} & \bfred{91.2\%} & \bfred{86.7\%} & \bfred{52.9\%} & \bfred{0.31} \\
        \textbf{QITE+iHVA} & \bfred{100.0\%} & 76.5\% & 80.3\% & 39.7\% & 0.64 \\
        \textbf{iHVA} & 98.5\% & 82.4\% & 75.6\% & 40.6\% & 4.46 \\
        \textbf{ma-QAOA} & 98.5\% & 51.5\% & 82.1\% & 24.1\% & 3.47 \\
        \textbf{HEA} & 98.5\% & 86.7\% & 81.4\% & 47.4\% & 2.35 \\
        \bottomrule
    \end{tabular}
    \label{tab:results}
\end{table*}

Table~\ref{tab:results} shows the feasibility and optimality within 5 trials, as well as the mean values of the metrics for
different methods. The results are averaged out of the 68 instances that we considered. For the feasibility and optimality columns,
an instance is recorded as feasible/optimal if it has obtained at least one feasible/optimal solution within 5 trials of VQA. Hence, the values shown in the table
are the percentage of instances (out of 68 instances) that achieved at least one feasible/optimal solution within 5 trials. For the rest of the columns, the
values are the mean of the metrics in the boxplot shown in Fig.~\ref{fig:results}.
It can be seen that the rescaled QITE+iHVA is triumphant in all metrics, and QITE achieves a drastically low mean optimality gap compared to VQE methods,
although the other metrics do not show much difference between QITE and VQE.

The reason why QITE achieved such a low mean optimality gap might be the nature of the optimization itself. In VQE, since the unitary gates of quantum circuits consist of sines and cosines,
the expectation function of the Hamiltonian is usually a product and sum of sines and cosines. This creates a nonconvex optimization problem with respect to the variational parameters, which is bound
to have many local minima. Nonconvex problems are notoriously difficult for local optimizers. On the other hand, QITE traces the (imaginary time) evolution of the Hamiltonian step-by-step with a very small time step.
For variational QITE, the parameters are updated using the McLachlan variational principle, which is different from optimizing based on
the gradient of $\langle\hat{H}\rangle$, therefore bypassing the nonconvex landscape of the expectation function~\cite{adapt-qite}. The challenge of QITE lies in the expressivity of the ansatz, i.e., whether it is able to estimate the imaginary-time-evolved quantum state
accurately at every time step.

\section{Discussions}
Other than the variational parameters, there are two important parameters in QITE that will affect the performance of the QITE optimization:
the total evolution time $\tau$ for the evolution $e^{-\tau\hat{H}}$ and the number of time steps $N_\tau$, related by
\begin{equation}
    \Delta\tau = \frac{\tau}{N_\tau},
    \label{eqn:time-step}
\end{equation}
where $\Delta\tau$ is the time step for the simulation. $\Delta\tau$ needs to be small enough to ensure to accuracy of the simulation.
In addition, the total evolution time $\tau$ should be long enough for the system to converge (refer to Fig.~\ref{fig:qite-individual}).
Assume that $\Delta\tau$ is fixed at a small value, more time steps $N_\tau$ are required to simulate the system for a longer evolution time $\tau$.
In practice, $N_\tau$ corresponds to the number of times that Eq.~(\ref{eqn:lse}) is solved (equivalent to the number of iterations in classical optimizers).
Solving Eq.~(\ref{eqn:lse}) requires simulating the quantum circuit to obtain the information about the trial state, as well as its gradient,
and then computing the inverse of $M$ to solve the linear system of equations.
This step is computationally expensive when the circuits are simulated classically.
Therefore, it is important to have a trade-off between $\tau$ and $N_\tau$ to ensure the evolution is long enough while keeping $N_\tau$ as small as possible.

Another factor that affects the performance of the evolution is the spectral norm of the problem Hamiltonian, defined as
\begin{equation}
    \lVert \hat{H}\rVert = \max_i |\lambda_i|,
\end{equation}
which is the eigenvalue of the Hamiltonian with the largest magnitude.
The spectral norm affects the rate at which the state evolves, i.e., a larger spectral norm causes
the quantum state to change faster throughout the evolution, resulting in the need for a smaller time step to track the evolution numerically.
This then leads to the rescaling of the problem Hamiltonian to ensure that its spectral norm is not too large for the evolution to be properly
simulated~\cite{rescaling}. Hence, we employ the rescaled Hamiltonian for QITE+iHVA:
\begin{equation}
    \hat{H}' = \frac{\hat{H}}{d},
    \label{eqn:ham-scale}
\end{equation}
where $d>1$ is a scalar. By setting $d=\lVert\hat{H}\rVert$, the coefficients can be restricted to the range $[-1,1]$.
As a consequence of scaling the Hamiltonian coefficients, the minimum energy of $\hat{H}$ will also be scaled by $1/d$, $E'_{\text{min}} = E_{\text{min}}/d$,
and hence can be retrieved by multiplying $d$ after solving the scaled Hamiltonian.

Scaling down the Hamiltonian coefficients is effectively equivalent to simulating evolution with a larger time step.
This can be easily seen by doing a substitution of Eq.~(\ref{eqn:ham-scale}) to (\ref{eqn:vi}), so the vector $V$ becomes $V/d$.
Propagating the scaled vector to Eq.~(\ref{eqn:euler}) gives 
\begin{equation}
    \bm{\theta}(\tau_0+\Delta\tau) = \bm{\theta}(\tau_0) + \frac{\dot{\bm{\theta}}}{d}\Delta\tau.
    \label{eqn:scaled-euler}
\end{equation}
Now we can afford to set a larger $\Delta\tau = d\Delta\tau'$ to amount for the same magnitude of parameters update as in Eq.~(\ref{eqn:euler}).
According to Eq.~(\ref{eqn:time-step}), if $\tau$ stays the same, a larger $\Delta\tau$ leads to a smaller number of time steps $N_\tau$ required for simulation,
Consequently, the simulation cost is reduced by scaling down the Hamiltonian coefficients. However, scaling down the coefficients does not always guarantee
convergence to the desired solution, which is shown in the results.

\begin{figure*}
    \centering
	\begin{subfigure}[t]{0.31\textwidth}
		\centering
		\sidecaption{subfig:a2}
		\raisebox{-\height}{\includegraphics[width=\textwidth]{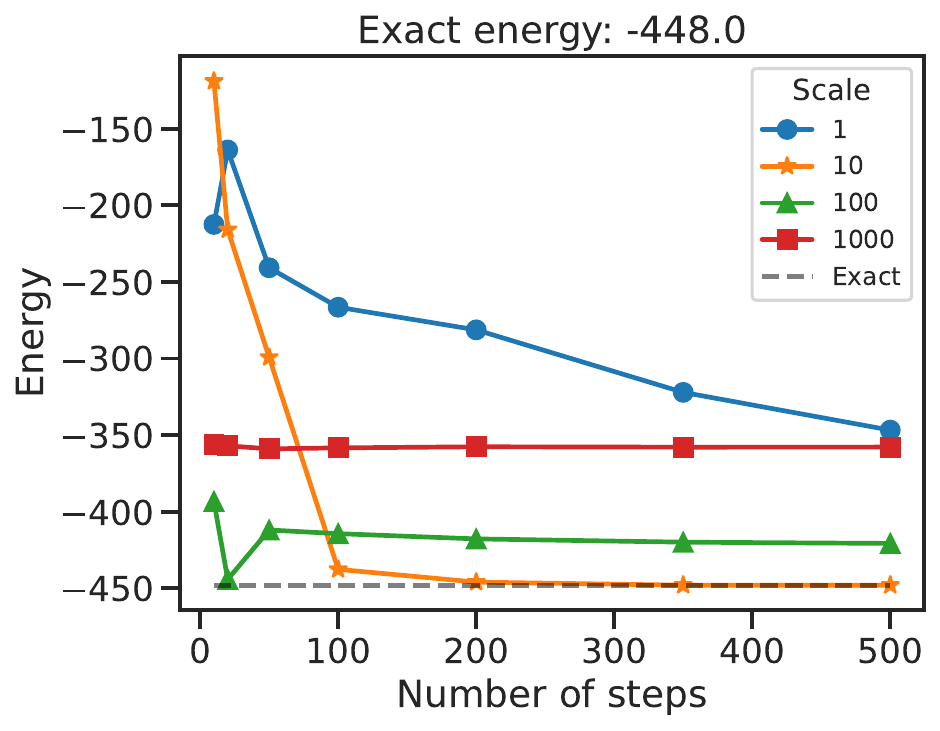}}
	\end{subfigure}
    \hfill
    \begin{subfigure}[t]{0.31\textwidth}
		\centering
		\sidecaption{subfig:b2}
		\raisebox{-\height}{\includegraphics[width=\textwidth]{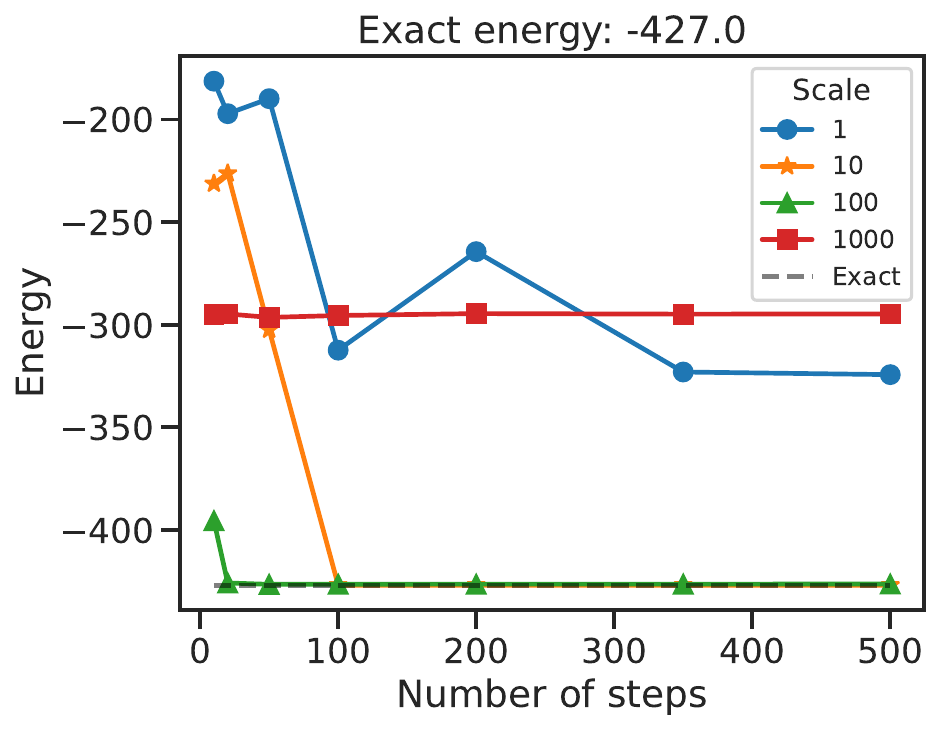}}
	\end{subfigure}
    \hfill
    \begin{subfigure}[t]{0.31\textwidth}
		\centering
		\sidecaption{subfig:c2}
		\raisebox{-\height}{\includegraphics[width=\textwidth]{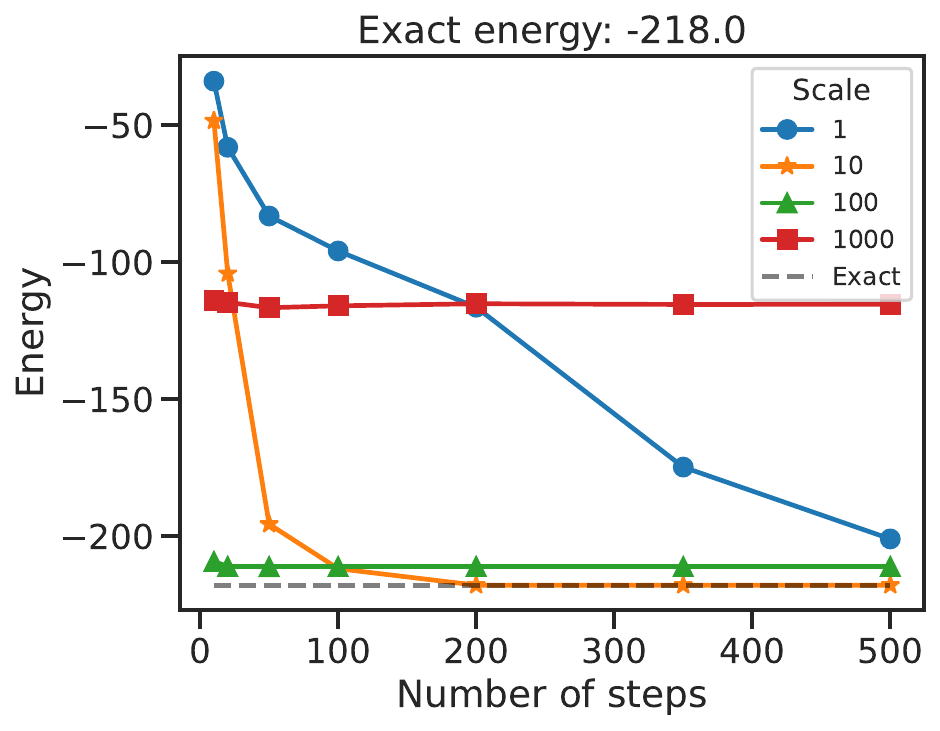}}
	\end{subfigure}
    \begin{subfigure}[t]{0.31\textwidth}
		\centering
		\sidecaption{subfig:d2}
		\raisebox{-\height}{\includegraphics[width=\textwidth]{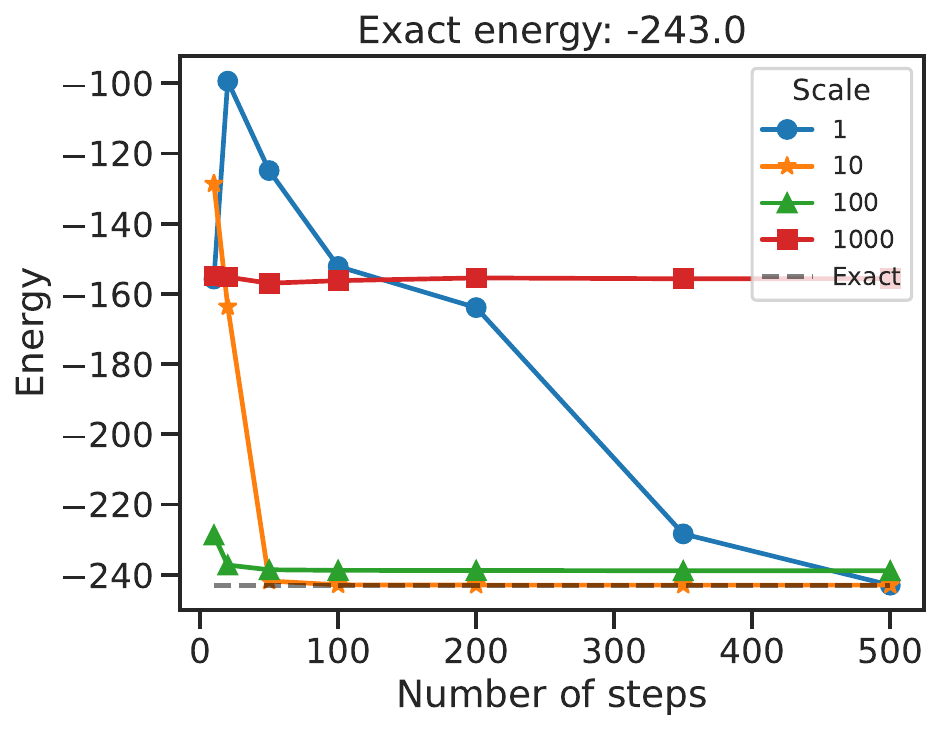}}
	\end{subfigure}
    \hfill
    \begin{subfigure}[t]{0.31\textwidth}
		\centering
		\sidecaption{subfig:e2}
		\raisebox{-\height}{\includegraphics[width=\textwidth]{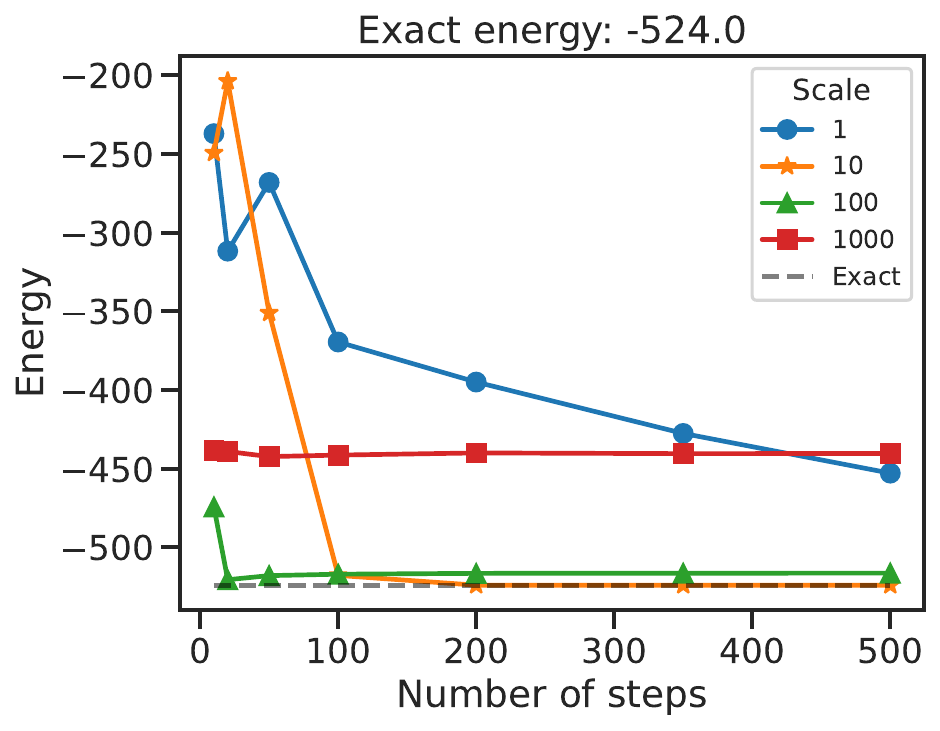}}
	\end{subfigure}
    \hfill
    \begin{subfigure}[t]{0.31\textwidth}
		\centering
		\sidecaption{subfig:f2}
		\raisebox{-\height}{\includegraphics[width=\textwidth]{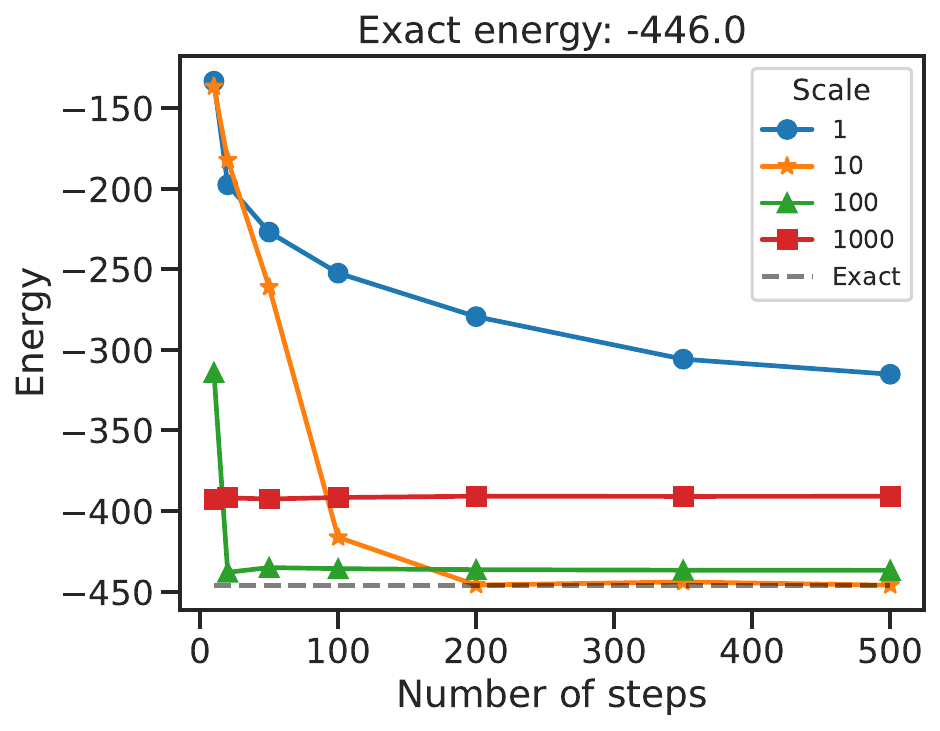}}
	\end{subfigure}
    \caption{Effect of scaling the Hamiltonian coefficients by $1/d$, with $d=1,10,100,\text{ and }1000$. Each plot shows the results for different Max-cut instances %
    (converted from the QUBO of MKP). %
    Each point in the plot shows the lowest energy attained after different number of steps $N_\tau$ of QITE. The dashed grey line shows the minimum energy %
    (solution) of the given instance.}
    \label{fig:scaling}
\end{figure*}

Fig.~\ref{fig:scaling} shows the variation of the lowest Max-Cut energy obtained against the number of steps used to simulate QITE+iHVA, with the Hamiltonian
coefficients scaled by different values. The Max-Cut energy originates from
\begin{equation}
    \langle \hat{H}_{\text{MaxCut}}\rangle = -\sum_{(i,j)}\expval{Z_iZ_j} + \frac{m}{2}.
\end{equation}
The first term in the RHS is often known as the \emph{energy} of the Hamiltonian $\hat{H}_{\text{MaxCut}}$ and the second term is a constant offset ($m$ is the number of edges in the MaxCut graph).
Each plot in Fig.~\ref{fig:scaling} shows the results for one MKP instance
(recall that we converted MKP to Max-Cut to adapt iHVA). Each point in a plot is the lowest energy obtained after
the convergence of QITE+iHVA by using different number of time steps. We use the total evolution time $\tau=10$ for the simulation, varying the number of steps $N_\tau$.
For $d=1$ (original Hamiltonian), the energy decreases in general with increase in $N_\tau$, but only (d) was able to
achieve the minimum energy of the Hamiltonian, i.e. the optimal solution, at $N_\tau = 500$. Others will need a larger number of time steps to achieve the minimum energy.
On the scale of $d=10$, the minimum energy is achieved between $N_\tau=50$ and $N_\tau=200$, which is significantly fewer than without scaling.
On the scale of $d=100$, we can see that although the energy hits its minimum as soon as $N_\tau = 50$,
the instances in (c), (d), (e), and (f) failed to achieve the exact energy, even increasing the number of time steps.
For $d=1000$, the system is unable to search the minimum energy at all and levels off as $N_\tau$ increases. This might be due to the long total evolution time used.
Fewer $N_\tau$ means that fewer iterations are needed for the evolution to converge to the minimum energy (the solution), thus saving the optimization cost for QITE.
Here, we can see that the best choices for the scales are $d=10$ for the best performance in terms of finding the solution, and $d=100$ for the best optimization cost
while keeping the solution sub-optimal.

\section{Conclusion}
We propose a framework (which we call QITE+iHVA) that converts a general QUBO instance into an equivalent Max-Cut instance in order to use the
Max-Cut-tailored iHVA ansatz. We then apply variational QITE to find the ground state (minimum energy) of the problem Hamiltonian.
The parameters are updated using the McLachlan variational principle, which requires solving a system of linear equations
at each update step. We found that QITE+iHVA achieved significantly smaller optimality gaps (solutions much closer to optimal) than all tested VQE methods
(including iHVA without QITE, ma-QAOA, and HEA) on our MKP instances.
The ability of QITE to obtain good solutions might be attributed to the way the parameters are updated, which avoids nonconvex optimization that is usually done in
conventional VQE methods.

However, the optimization of VarQITE is computationally expensive when simulated classically, as it involves computing the trial state and its gradients, as well
as solving a system of linear equations at every update step. Therefore, deciding the time step for the evolution is important to reduce the optimization cost.
We found that by scaling the Hamiltonian coefficients, we effectively allow larger time steps to update the parameters with the same amount of magnitude.
Through the experiments, we found that suitable scaling of the Hamiltonian can also lead to better solutions.

The main future work is to cope with constrained problems with larger size, including optimizing the implementation of the McLachlan variational principle
to speed up the computation time. As problems become larger, deeper circuits might be needed for better expressivity to represent the imaginary-time-evolved
state at each time step. On the other hand, heuristics that cleverly determine the scale of the Hamiltonian are required, instead of finding the scales
empirically.

\appendix
The algorithm for the conversion of a QUBO instance to its equivalent Max-Cut instance is mentioned in~\cite{qubo2maxcut} and briefly in the
appendix of~\cite{shrinking-vrp}.
The main idea of the algorithm is to first convert the binary variables $x_i\in\{0,1\}$ in QUBO into the spin variables $s_i\in\{-1,1\}$.
Then an additional variable $s_0$ is introduced to make the linear terms quadratic, so that the entire expression contains only quadratic terms,
which is exactly a Max-Cut problem.
The edge weights $w_{ij}$ for the Max-Cut graph can then be determined by rearranging the coefficients. 

\begin{algorithm}[H]
    \caption{QUBO to Max-Cut}
    \textbf{Input:} QUBO problem: $\sum_{i,j>i}^n q_{ij}x_ix_j + \sum_{i=1}^n l_ix_i,\ \forall i,j\in\{1,...,n\}$
    \begin{algorithmic}[1]
        \State Assign the linear coefficients into the QUBO matrix: $q_{ii} := l_i$
        \State Create a graph $G=(V,E)$ with $n+1$ vertices, $V\in\{0,1,...,n\}$.        
        \State Assign weights for edge $(0,i)$:
        \Statex $w_{0i} := \sum_{j=1}^n q_{ij} + q_{ji}$.
        \State Assign weights for all other edges:
        \Statex $w_{ij} := q_{ij} + q_{ji}$.
    \end{algorithmic}
    \textbf{Output:} Weighted graph $G'=(V,E,w)$.
    \label{alg:qubo-to-maxcut}
\end{algorithm}

After solving the Max-Cut problem for $G'$, the solution for the QUBO is reconstructed by 
setting $x_i = 1$ if the edge $(0,i)$ of $G'$ is cut, or else $x_i = 0$. 
\newpage
\bibliography{qaoaref,qiteref,ccop}
\bibliographystyle{ieeetr}
\end{document}